\documentclass[12pt]{article}
\usepackage{graphicx}
\usepackage{amsmath}
\usepackage{amssymb}
\usepackage{amsthm}
\theoremstyle{plain}

\newtheorem{lem}{Lemma}
\newtheorem{prop}{Proposition}

\theoremstyle{definition}

\theoremstyle{remark}
\newtheorem{rem}{Remark}
\author{Shawn Westmoreland}
\title{On the principle of relativity and electrostatics in a flat spacetime with a compact spatial dimension}
\date{\emph{Dept. of Mathematics, Kansas State University}}
\begin{document}
\maketitle
\begin{abstract}We study a flat spacetime with a compact spatial dimension and dispute the claim of Bansal, Laing, and Sriharan that a local experiment can determine whether or not an inertial observer is ``privileged" in this spacetime.
\end{abstract}
\section{Introduction} Flat spacetimes are not as sophisticated or realistic as their curved counterparts, but they are still worth studying as they can provide insight to general relativity. For example, there are the Milne and Rindler universes (see \cite{1}), and, in a quite different flavor, there is the ``twin paradox on compact space" (see e.g. \cite{2}-\cite{6}).

The paper by Bansal, Laing, and Sriharan \cite{6} concerns a flat spacetime $M$ with topology $S^1\times\mathbb{R}^{2,1}$. This spacetime can be constructed from standard Minkowski spacetime $\mathbb{R}^{3,1}$ by declaring points with coordinates $(x,y,z,t)$ \emph{equivalent} to points with coordinates $(x+nL,y,z,t)$, where $L>0$ measures  the length of the compact ``circular" spatial dimension and $n\in\mathbb{Z}$. The resulting manifold $M$ inherits the flat Minkowskian metric $ds^2=dx^2+dy^2+dz^2-dt^2$ from $\mathbb{R}^{3,1}$. Each event (or point) in $M$ corresponds to an equivalence class of points $[(x+nL,y,z,t)]_{n\in\mathbb{Z}}$ in $\mathbb{R}^{3,1}$. Hence, each point in $M$ can be uniquely expressed by coordinates $(x,y,z,t)$ such that $0\leq x< L$. In other words, one has a quotient map which projects each point $(x,y,z,t)$ of $\mathbb{R}^{3,1}$ to the point of $M$ given by coordinates $(x\mod L,y,z,t)$. The worldline of an inertial observer in $M$ can be regarded as the projection of an inertial observer in $\mathbb{R}^{3,1}$

The equivalence relation $(x,y,z,t)\sim(x+nL,y,z,t)$, used above to describe the topology of $M$, is coordinate dependent. The Lorentz transformation given by the matrix 
\begin{displaymath}
[\Lambda^{\nu'}_{\mu}]=
\left[ \begin{array}{rrrr}
(\gamma-1)(u^1)^2+1 & (\gamma-1)u^1u^2 & (\gamma-1)u^1u^3 & -\beta\gamma u^1 \\
(\gamma-1)u^1u^2 & (\gamma-1)(u^2)^2+1 & (\gamma-1)u^2u^3&-\beta\gamma u^2\\
(\gamma-1)u^1u^3&(\gamma-1)u^2u^3&(\gamma-1)(u^3)^2+1&-\beta\gamma u^3\\
-\beta\gamma u^1&-\beta\gamma u^2&-\beta\gamma u^3&\gamma
\end{array}\right],
\end{displaymath}
where $\gamma:=(1-\beta^2)^{-\frac{1}{2}}$, describes the transformation from the ``unprimed" frame $(x,y,z,t)$ to a ``primed" frame $(x',y',z',t')$ in $\mathbb{R}^{3,1}$ which moves at a uniform velocity $\beta\vec{u}$, with $\vec{u}=(u^1,u^2,u^3)$ such that $|\vec{u}|^2:=(u^1)^2+(u^2)^2+(u^3)^2=1$, relative to the unprimed frame (see e.g. \cite{7}, page 69).  For such an observer, the equivalence relation for describing the topology of $M$ becomes, for any $n\in\mathbb{Z}$,
$$(x',y',z',t')\sim(x'+n((\gamma-1)(u^1)^2+1)L,y'+n(\gamma-1)u^1u^2 L,z'+n(\gamma-1)u^1u^3L,t'-n\beta\gamma u^1L).$$ Whence, a general boosted observer identifies points not only across space, but also across time (if $-\beta\gamma u^1 L\neq0$). If this ``time shift" $-\beta\gamma u^1L$ is nonzero, then there exist two separated spatial positions, which would normally be regarded as simultaneous by the Einstein convention, that are assigned two different times by this observer. Hence, such an observer (or more precisely, the projection of such an observer down to $M$) cannot set up a ``global" system of Einstein-synchronized clocks encompassing the entire universe. On the other hand, if $-\beta\gamma u^1L=0$, then the observer in Minkowski spacetime identifies (what she regards as) simultaneous spatial positions through space only, preserving the assignment of a unique time to each event. Hence, such an observer has the special privilege of being able to set up a global system of Einstein-synchronized clocks in $M$. We will call these specially privileged observers, \textbf{privileged}. The upshot is that $M$ does not have a unique privileged frame but instead has infinitely many privileged frames (those described by setting either $\beta=0$ or $u^1=0$, so that $-\beta\gamma u^1L=0$). These privileged frames exhibit inertial motions with respect to each other. This is unlike the flat spacetime with topology $S^1\times\mathbb{R}^1$ studied in \cite{2}. Flat $S^1\times\mathbb{R}^1$ spacetime has a \emph{unique} privileged frame, which by virtue of its uniqueness can be regarded as an ``absolute rest frame."

\begin{figure}[!p] 
\centering
   \includegraphics[width=4in]{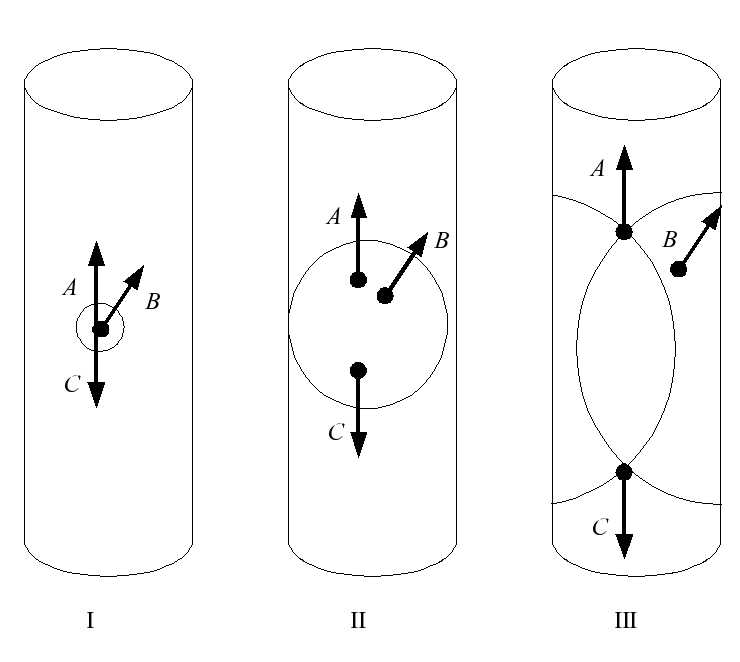}
\caption{With one spatial dimension suppressed, three ``snapshots" of the ``universe" are shown. I). Light flash is initiated. II). Light flash expands into all directions at the speed of light. III). Observers $A$ and $C$ receive the flash from two directions at the same time (according to their own respective observations). Observer $B$ has already received a flash from one direction and will soon receive a flash from the other. The conclusion is that observers $A$ and $C$ are privileged, but observer $B$ is not.} 
\label{cylinders3}  
\end{figure}

Here is an elementary physical argument for the non-uniqueness of privileged frames in $M$. An inertial observer can determine whether or not she is privileged  by sending out a flash of light from her position. This flash of light will spread out in all directions, including the compact dimension, at the speed of light.  By looking along the compact dimension, the observer will eventually be able to see the light flash that she had previously initiated. If she is moving along a non-compact dimension, but \emph{not} moving along the compact dimension,\footnote{That is, \emph{with respect to a given privileged frame}.} then she will see the light flash in both directions along the compact dimension at the same time, and hence she will discover that she is privileged. If she \emph{is} moving along the compact dimension, then she  will see her previous light flash in one direction before seeing it in the opposite direction, and hence she will learn that she is not privileged. (See figure~\ref{cylinders3}). 

This raises an issue. By using simple physical experiments involving flashes of light, inertial observers can learn whether or not they are privileged. Hence the principle of relativity, strictly applied, breaks down in this model.\footnote{We take the ``strict" version of the principle of relativity to state that the laws of physics do not distinguish between two inertial reference frames. In the words of Einstein, written in 1905, ``the laws by which the states of physical systems undergo change are not affected, whether these changes of state be referred to the one or the other of two systems of coordinates in uniform translatory motion." (See e.g., \cite{11}, page 41.) Hence, according to the principle of relativity, there is no physical experiment an observer can perform in order to determine whether or not she is privileged.}

However, the principle of relativity does hold \emph{locally} in $M$, as we show in section 3 after the locality concept is sufficiently clarified.\footnote{Note that ``locally" is all that really matters since in general relativity, the rules of special relativity need only hold locally (see also \cite{2}). To say that the principle of relativity holds locally is to say that the \emph{local} laws of physics do not distinguish between two inertial reference frames. We define the local laws of physics to be the laws of physics that are verified by local experiments. The notion of a local experiment is discussed in section 3.} Thus, inertial observers in $M$ cannot use \emph{local} experiments to determine whether or not they are privileged.

The authors of \cite{6} propose, to the contrary, that inertial observers in $M$ \emph{can} use a local experiment (involving electrostatics) to determine whether or not they are privileged. In section 3 we will explain why their experiment is not really local, but first we need to describe the electrostatic field in $M$. Our analysis corrects errors made by \cite{6} on the matter.

\section{Electrostatics in flat $S^1\times\mathbb{R}^{2,1}$}

In \cite{6}, the analysis is implicitly constrained to the case where the ``primed" frame exhibits inertial motion with respect to the ``unprimed" frame at a velocity $\beta\vec{u}$, where $\vec{u}=(u^1,u^2,u^3)=(1,0,0)$. In such a frame, the equivalence relation describing the topology of $M$ becomes, for any $n\in\mathbb{Z}$,
$$(x',y',z',t')\sim(x'+n\gamma L,y',z',t'-n\beta\gamma L).$$ Note that in this frame, the length of the compact spatial dimension is effectively $L_{eff}=\gamma L$. This is somewhat larger than the value found by a privileged observer.\footnote{Note that for a general boosted observer, having velocity $\beta\vec{u}$ with respect to the unprimed frame, the effective size of the compact dimension is
$L_{eff}=L\sqrt{(\gamma-1)^2(u^1)^2+2(\gamma-1)(u^1)^2+1}$. The physical meaning of $L_{eff}$ is that an inertial observer in $M$ cannot extend her coordinate system farther than $L_{eff}$ in the compact direction, if her coordinate system synchronizes clocks by the Einstein procedure. Note that $L_{eff}\geq L$, with equality holding if and only if one is in a privileged frame.}
Assume that at the spatial origin of such a frame, there sits a stationary point-charge. It has existed there for all past eternity and will continue to exist there for all future eternity. Its electric charge is, always has been, and always will be, an absolute nonzero constant $q$. There are no other electric charges or magnetic fields in $M$.

Since $M$ has the topology $S^1\times \mathbb{R}^{2,1}$, one expects that the electrostatic field of such a point-charge should deviate from Coulomb's inverse-square law. Indeed, the $x'$-component of the electrostatic field generated by the charge should vanish at the ``antipodal" point on the $x'$-axis with coordinate $L_{eff}/2$. So the electrostatic field in $M$ should decrease \emph{faster} than the inverse-square of the distance along the $x'$-axis.

As explained in \cite{6}, the electrostatic field at the spatial point $(x,y,z)$ in $M$ in such a reference frame is given by\footnote{We are neglecting interactions between the metric and the electromagnetic field and mass so that we can assume the spacetime is flat.} $$\stackrel{\rightarrow}{E}(x,y,z)=\frac{q}{4\pi\epsilon_0}\sum_{n=-\infty}^\infty \frac{(x+nL)\hat{x}+y\hat{y}+z\hat{z}}{[(x+nL)^2+y^2+z^2]^\frac{3}{2}},$$ where we have dropped the ``primes" and written $L$ in place of $L_{eff}$ for simplicity. Consider the $x$-component of the electric field along the $x$-axis (where $y$ and $z$ vanish). It is given by
$$E_{\hat{x}}(x)=\frac{q}{4\pi\epsilon_0}\sum_{n=-\infty}^\infty\frac{x+nL}{[(x+nL)^2]^\frac{3}{2}}.$$ This function is periodic with period $L$, as one would expect since the $x$-dimension is periodic with length $L$ (for an analytical proof, see proposition 1 in the appendix). Note that, if $0<x<L$,
\begin{eqnarray}
\sum_{n=-\infty}^\infty\frac{x+nL}{[(x+nL)^2]^\frac{3}{2}}&=&\frac{1}{x^2}+\sum_{n=1}^\infty\left(\frac{1}{(x+nL)^2}-\frac{1}{(x-nL)^2}\right) \nonumber \\
&=&\frac{1}{x^2}+\frac{1}{L^2}\sum_{n=1}^\infty\left(\frac{1}{(x/L+n)^2}-\frac{1}{(x/L-n)^2}\right). \nonumber
\end{eqnarray}
Defining a function $\varphi$ through the series$$\varphi(z):=\sum_{n=1}^\infty\left(\frac{1}{(z+n)^2}-\frac{1}{(z-n)^2}\right),$$which converges absolutely for $z\in \mathbb{C}\setminus\mathbb{Z}$ (proposition 2 in the appendix), we can write, for $0<x<L$,
$$\frac{4\pi\epsilon_0}{q}E_{\hat{x}}(x)=\frac{1}{x^2}+\frac{1}{L^2}\varphi\left(\frac{x}{L}\right).$$ Note that $\varphi(\frac{1}{2})=-4$. Hence, $E_{\hat{x}}(L/2)=0$, as expected.

Moreover, one verifies that the series defining $\varphi(z)$ converges uniformly over compact subsets of the unit disc $\mathbb{D}=\{z\in\mathbb{C}:\ |z|<1\}$ (lemma 1 in the appendix). It follows (see \cite{9}, page 214), that $\varphi$ is holomorphic on $\mathbb{D}$ and all the derivatives of $\varphi$ can be calculated by differentiating behind the summation symbol. One shows by mathematical induction that $$\varphi^{(k)}(z)=(-1)^{k}(k+1)!\sum_{n=1}^\infty\left[(z+n)^{-2-k}-(z-n)^{-2-k}\right].$$ Hence $\varphi^{(k)}(0)$ vanishes if $k$ is even and $\varphi^{(k)}(0)=-(k+1)!2\zeta(k+2)$ if $k$ is odd, where $\zeta(s):=\sum_{n=1}^\infty n^{-s}$ is the Riemann zeta function (see e.g., \cite{10}, page 212). We therefore have a power series expansion
$$\varphi(z)=-4\sum_{n=0}^\infty(n+1)\zeta(2n+3)z^{2n+1}.$$
It can be verified that the radius of convergence for this power series is 1 (proposition 3 in the appendix). In other words, the power series is valid on $\mathbb{D}$.

Consequently, for $0<x<L$,
\begin{eqnarray}
\frac{4\pi\epsilon_0}{q}E_{\hat{x}}(x)&=&\frac{1}{x^2}-\frac{4}{L^2}\sum_{n=0}^\infty(n+1)\zeta(2n+3)(x/L)^{2n+1} \nonumber \\
&=&\frac{1}{x^2}-\frac{4}{L^2}\left(\zeta(3)(x/L)+\mathcal{O}(x^3/L^3) \right).\nonumber
\end{eqnarray} So we get that the first-order correction term in $M$ to the usual Coloumb law is $-\frac{q\zeta(3)}{\pi\epsilon_0L^3}x$.  The 
absolute fractional difference from the usual electrostatic field (along the $x$-axis) is therefore approximated by
$$\left|\frac{\Delta E}{E}\right|\approx 4\zeta(3)\left(\frac{x}{L}\right)^3.$$

The fact that $\varphi$ is an odd function with $\varphi(0)=0$, and not, as reported in \cite{6}, an even function which is constant to first-order, is anticipated by the following physical argument. 

Let $\varphi$ denote the correction function to the electrostatic field along the $x$-axis. Note that we are now using the name  ``$\varphi$" somewhat abusively, since previously $(q/(4\pi\epsilon_0 L^2))\varphi(x/L)$ is the correction to the electrostatic field along the $x$-axis where $0< x<L$. For the present argument, we are taking $L=1$ and
$$\varphi(x):=E_{\hat{x}}(x)-\frac{q}{4\pi\epsilon_0}\frac{x}{|x|^3},$$where $x\in (-1,1)\setminus\{0\}$. Physically, $\varphi(a)$ can be interpreted as the strength of an electrostatic field at the point $x=a\notin\mathbb{Z}$ on the $x$-axis (in $\mathbb{R}^3$) generated by an infinite sequence of point-charges, each of charge $q$, individually glued down (so that they don't move) onto the $x$-axis at the coordinates $x=\pm1, \pm2,\pm3,...$. By symmetry, the electrostatic fields of the individual charges cancel out near $x=0$. In other words, $\lim_{x\rightarrow0}\varphi(x)=0$. Moreover, since the electrostatic field on the $x$-axis at the points $x=a$ and $x=-a$ (with $x\in (-1,1)\setminus\{0\}$) should have the same magnitude but opposite direction, one expects that $\varphi(a)=-\varphi(-a)$. That is, $\varphi$ is an odd function.
\newpage
\section{A local experiment?}

In \cite{6} the claim is made that in the spacetime $M$ a ``local" experiment, involving the electrostatic field of a point-charge, can be used to determine whether or not an inertial observer is privileged. The idea is to note that the law for electrostatics in $M$ depends on the (effective) size of the compact dimension, as we have found in section 2. As noted in section 2, the compact dimension has length $L$ in a privileged frame but has a somewhat larger length in a non-privileged frame. The authors of \cite{6} conclude from these facts that an inertial observer in $M$ can determine (or at least estimate) the effective size of the compact dimension in their frame by making ``nearby" measurements of the electrostatic field strength generated by a stationary point-charge in their frame. The results found in section 2 support the idea that this can be done, at least in principle.\footnote{However, we have not addressed the subtleties that arise when the charge is transferred from being stationary in one reference frame to being stationary in another, which is what one might do in order to use this particular experiment to compare two reference frames. The analysis in section 2 is restricted to the case where the charge is stationary in one frame for all eternity. However, our main concern is not with the finer details of electromagnetism in $M$ but rather the more immediate question of whether the experiment proposed by \cite{6} is truly local, even if it is done in only one frame.} However, we do not agree with the claim that this is a truly ``local" experiment. The principle of relativity, at least in general relativity, says that the \emph{local} laws of physics do not  distinguish between two inertial observers. Hence, there is no \emph{local} experiment that an observer can perform in order to determine whether or not she is privileged. Therefore, the ``local" experiment proposed by \cite{6} cannot be called ``local" if we wish to retain the principle of relativity. Yet, a dogmatic commitment to the principle of relativity does nothing to explain why the experiment is not a local one. At a more basic level, one can argue as follows.

We start by investigating the meaning of the word 
``local." In the present context one can distinguish between two notions of locality: 
\textbf{geometrical locality} and \textbf{topological locality} (compare \cite{3}). We shall say that 
a physical system is local in the geometrical sense if it exists wholly within ``an infinitesimal 
region" of the spacetime manifold. On the other hand, we shall say that a physical system is 
local in the topological sense if it exists wholly within a region we call ``a patch"\footnote{A ``patch" of a flat spacetime is an open region isometric to an open region of the Minkowski spacetime with the same dimension. A flat, differentiable spacetime is locally isometric to a Minkowski spacetime (e.g. \cite{12}, page 208) and hence is a union of ``patches."} of the 
manifold. Since an infinitesimal region (whatever that is) can be contained 
within a single patch of $M$, it follows that the topological notion of 
locality generalizes the geometrical notion.\footnote{Warning: our notion of ``topological" locality is only useful in flat regions of spacetime.} Thus, if a system is not local in 
the topological sense, it cannot be local in the geometrical sense either. 
It is sufficient for our purposes to note that a properly local experiment
necessarily involves only a physical system which can be completely described 
within one patch of the spacetime manifold.

We can apply this notion to see why the experiment described in the introduction of this note, the experiment involving a light flash, is not local. This thought experiment concerns a physical system involving light signals which traverse the entire compact dimension of $M$. Hence, at least \emph{two} patches of $M$ would be needed in order to completely describe the experiment. In a certain na\"ive sense, this experiment only requires the observer to make a few local time measurements on his or her worldline; whence we notice that even if an experiment wholly relies on measurements which na\"ively appear to be ``local," it does not follow that the whole experiment itself really \emph{is} local. 

The same argument carries over to the ``local" thought experiment proposed by \cite{6}. Their experiment only requires the observer to make apparently ``local" (or ``nearby") measurements of an electrostatic field, but this does not mean that the experiment is truly local. Indeed, a more \emph{complete} description of the experiment would account for the underlying physical systems which influence each measurement. For example, consider  the question: what produces the ``local" electrostatic field and what determines its structure? The producer of the field is the point-charge source (in the sense that, in this thought experiment, there is no field without the point-charge), and the structure of the field (at any point where it is defined) depends on ``global" topological features of the manifold (a fact which is not overlooked in \cite{6}). Hence, the experiment proposed by \cite{6}, if described more completely, actually involves a physical system which includes the point-charge source together with an electrostatic field generated by the charge, which wraps all the way around the compact dimension. This physical system cannot be described within a single patch of the manifold. Therefore, the experiment is not local.

Indeed, the principle of relativity \emph{must} hold locally in $M$. A local experiment is a physical system which can be completely contained in a single patch of $M$. A patch of $M$ is essentially indistinguishable from a region of Minkowski spacetime $\mathbb{R}^{3,1}$, in which the principle of relativity certainly holds. This argument is independent of the topology of $M$. (See also \cite{2}.)

\section{Appendix}
The purpose of this appendix is to fill in the mathematical gaps left in section 2. Throughout, $L$ is a positive constant.
\begin{prop}The function $$\frac{4\pi\epsilon_0}{q}E_{\hat{x}}(x)=\sum_{n=-\infty}^\infty\frac{x+nL}{[(x+nL)^2]^\frac{3}{2}}$$ is absolutely convergent for $x\in\mathbb{R}\setminus\{nL:\ n\in\mathbb{Z}\}$ and periodic with period $L$. 
\end{prop}
\begin{proof} Let $x\in\mathbb{R}\setminus \{nL:\ n\in\mathbb{Z}\}$. According to the interpretation of a doubly infinite sum (e.g., \cite{10}, page 184), we have
\begin{eqnarray}\sum_{n=-\infty}^\infty\frac{x+nL}{[(x+nL)^2]^\frac{3}{2}}&:=& \sum_{n=0}^\infty\frac{x+nL}{[(x+nL)^2]^\frac{3}{2}}+\sum_{n=1}^\infty\frac{x-nL}{[(x-nL)^2]^\frac{3}{2}} \nonumber \\
&=&\sum_{n=0}^\infty\frac{x+nL}{|x+nL|^3}+\sum_{n=1}^\infty\frac{x-nL}{|x-nL|^3}.
\nonumber
\end{eqnarray}
Note that $|(x+nL)|x+nL|^{-3}|=(x+nL)^{-2}$ and  $|(x-nL)|x-nL|^{-3}|=(x-nL)^{-2}$. For $N>|x/L|$, one verifies that both $\sum_{n=N}^\infty(x+nL)^{-2}$ and $\sum_{n=N}^\infty(x-nL)^{-2}$ converge by the ``integral test" (see \cite{8}, page 139). Hence, it follows that $\sum_{n=-\infty}^\infty\frac{x+nL}{[(x+nL)^2]^\frac{3}{2}}$ converges absolutely. Moreover, we can write
$$\sum_{n=-\infty}^\infty\frac{x+nL}{[(x+nL)^2]^\frac{3}{2}}=\frac{x}{|x|^3}+\sum_{n=1}^\infty\left(\frac{x+nL}{|x+nL|^3}+\frac{x-nL}{|x-nL|^3}\right).$$

Note that
\begin{eqnarray}
\frac{4\pi\epsilon_0}{q}E_{\hat{x}}(x)&=&\sum_{n=0}^\infty\frac{x+nL}{|x+nL|^3}+\sum_{n=1}^\infty\frac{x-nL}{|x-nL|^3}\nonumber \\ &=&\frac{x}{|x|^3}+\sum_{n=1}^\infty\frac{x+nL}{|x+nL|^3}+\sum_{n=1}^\infty\frac{x-nL}{|x-nL|^3},\nonumber
\end{eqnarray}and
\begin{eqnarray}
\frac{4\pi\epsilon_0}{q}E_{\hat{x}}(x+L)&=&\sum_{n=0}^\infty\frac{x+(1+n)L}{|x+(1+n)L|^3}+\sum_{n=1}^\infty\frac{x+(1-n)L}{|x+(1-n)L|^3}\nonumber \\ &=&
\sum_{n=1}^\infty\frac{x+nL}{|x+nL|^3}+\frac{x}{|x|^3}+\sum_{n=1}^\infty\frac{x-nL}{|x-nL|^3}.\nonumber
\end{eqnarray}
Hence, $E_{\hat{x}}(x)=E_{\hat{x}}(x+L)$ and so the function is periodic with period $L$.
\end{proof}
\begin{prop}The series $$\varphi(z)=\sum_{n=1}^\infty\left(\frac{1}{(z+n)^2}-\frac{1}{(z-n)^2}\right)$$ converges absolutely for $z\in\mathbb{C}\setminus\mathbb{Z}$.
\end{prop}
\begin{proof}For $z\in\mathbb{C}\setminus\mathbb{Z}$ and $n>|z|$, we have that
\begin{eqnarray}
\left|\frac{1}{(z+n)^2}-\frac{1}{(z-n)^2}\right|&=&\left|\frac{4zn}{(n^2-z^2)^2}\right| \nonumber \\
&\leq&\frac{4|z|n}{(n^2-|z|^2)^2}.\nonumber
\end{eqnarray}
Choosing $N>|z|$, one finds that the series $\sum_{n=N}^\infty \frac{4|z|n}{(n^2-|z|^2)^2}$ converges by the integral test. 
Whence, the series $\sum_{n=1}^\infty\left(\frac{1}{(z+n)^2}-\frac{1}{(z-n)^2}\right)$ converges absolutely for $z\in\mathbb{C}\setminus\mathbb{Z}$, by the so-called ``comparison test" (\cite{8}, page 60).
\end{proof}
The remainder of this appendix is devoted to proving the following proposition.
\begin{prop}The function $$\varphi(z)=\sum_{n=1}^\infty\left(\frac{1}{(z+n)^2}-\frac{1}{(z-n)^2}\right)$$ has a power series representation 
$$\varphi(z)=-4\sum_{n=0}^\infty(n+1)\zeta(2n+3)z^{2n+1},$$ which converges on the unit disc $\mathbb{D}=\{z\in\mathbb{C}:\ |z|<1\}$.
\end{prop}
The proof of this will proceed from a few lemmas.
\begin{lem}The series $$\varphi(z)=\sum_{n=1}^\infty\left(\frac{1}{(z+n)^2}-\frac{1}{(z-n)^2}\right)$$ converges uniformly over compact subsets of $\mathbb{D}$. 
\end{lem}
\begin{proof} It suffices to show that the series converges uniformly on $\overline{D}(0;r):=\{z\in\mathbb{C}: |z|\leq r\}$, where $r<1$. We have that, for $z\in\mathbb{D}$, 
$$\left|\frac{1}{(z+n)^2}-\frac{1}{(z-n)^2}\right|=\frac{4|z|n}{|n^2-z^2|^2}.$$ Note that, for $|z|\leq r$, we have that $0<n^2-|z|^2\leq |n^2 - z^2|$. Hence, $0<(n^2-r^2)^2\leq |n^2-z^2|^2$, and we find that 
$$\left|\frac{1}{(z+n)^2}-\frac{1}{(z-n)^2}\right|=\frac{4|z|n}{|n^2-z^2|^2}\leq\frac{4rn}{(n^2-r^2)^2}.$$ The series $\sum_{n=1}^\infty \frac{4rn}{(n^2-r^2)^2}$ converges by the integral test. It now follows, by the ``Weierstrass $M$-test," (\cite{10}, page 37) that $$\varphi(z)=\sum_{n=1}^\infty\left(\frac{1}{(z+n)^2}-\frac{1}{(z-n)^2}\right)$$ converges uniformly over $\overline{D}(0;r)$.
\end{proof}

\begin{lem}For any $z\in\mathbb{D}$, $$\varphi^{(k)}(z)=(-1)^k(k+1)!\sum_{n=1}^\infty\left[(z+n)^{-2-k}-(z-n)^{-2-k}\right],$$ and hence 

\begin{displaymath}
\varphi^{(k)}(0)=\left\{ \begin{array}{ll}
0 & \textrm{if $k$ is even}\\
-(k+1)!2\zeta(k+2) & \textrm{if $k$ is odd.}
\end{array} \right.
\end{displaymath}
\end{lem}
\begin{proof} Since we have verified that $\varphi(z)=\sum_{n=1}^\infty \left[(z+n)^{-2}-(z-n)^{-2}\right]$ converges uniformly over compact subsets of $\mathbb{D}$ (lemma 1), we have that the sequence $\{\varphi_N (z)\}_{N=1}^\infty$, where $$\varphi_N(z):=\sum_{n=1}^N \left[(z+n)^{-2}-(z-n)^{-2}\right],$$ converges uniformly over compact subsets of $\mathbb{D}$ to $\varphi$. Note also that each $\varphi_N$ is holomorphic on $\mathbb{D}$.

Therefore, by theorem 10.28 and its corollary (page 214 in \cite{9}), we have that $\varphi$ is holomorphic on $\mathbb{D}$ and $\varphi_N^{(k)}\rightarrow\varphi^{(k)}$ uniformly, as $N\rightarrow \infty$, on compact subsets of $\mathbb{D}$. 

Since 
\begin{eqnarray}
\lim_{N\rightarrow\infty}\varphi_N^{(k)}&=&\lim_{N\rightarrow\infty}\frac{d^k}{dz^k}\sum_{n=1}^N \left[(z+n)^{-2}-(z-n)^{-2} \right]\nonumber \\ 
&=&\lim_{N\rightarrow\infty}\sum_{n=1}^N\frac{d^k}{dz^k}\left[(z+n)^{-2}-(z-n)^{-2}\right] \nonumber \\
&=&\sum_{n=1}^\infty\frac{d^k}{dz^k}\left[(z+n)^{-2}-(z-n)^{-2}\right],\nonumber
\end{eqnarray}
we have that the derivatives of $\varphi$ can all be calculated by differentiating behind the summation symbol.

To complete the proof of this lemma, we show that  $$\varphi^{(k)}(z)=(-1)^k(k+1)!\sum_{n=1}^\infty \left[(z+n)^{-2-k}-(z-n)^{-2-k}\right].$$ This formula can be established by mathematical induction. We have that $\varphi^{(0)}(z):=\varphi(z)$, and according to the formula, 
\begin{eqnarray}
\varphi^{(0)}(z)&=&(-1)^0(0+1)!\sum_{n=1}^\infty \left[(z+n)^{-2-0}-(z-n)^{-2-0}\right]\nonumber \\
&=&\sum_{n=1}^\infty\left[(z+n)^{-2}-(z-n)^{-2}\right]\nonumber \\
&=&\varphi(z).\nonumber
\end{eqnarray}

For induction, suppose $\varphi^{(m)}(z)=(-1)^m(m+1)!\sum_{n=1}^\infty \left[(z+n)^{-2-m}-(z-n)^{-2-m}\right].$ Then, since we can differentiate behind the summation symbol, 
\begin{eqnarray}
\varphi^{(m+1)}(z)&=&(-1)^m(m+1)!\sum_{n=1}^\infty \left[(-2-m)(z+n)^{-2-m-1}-(-2-m)(z-n)^{-2-m-1}\right]\nonumber \\
&=&(-1)^{m+1}(m+2)!\sum_{n=1}^\infty\left[(z+n)^{-2-(m+1)}-(z-n)^{-2-(m+1)}\right]. \nonumber
\end{eqnarray}
Thus, by induction, the lemma holds.
\end{proof}
\begin{rem}Rudin (page 214 in \cite{9}) remarks that for functions on the real line, it is possible for sequences of infinitely differentiable functions to converge uniformly to nowhere differentiable functions. This is the reason why we find it desirable to consider $\varphi$ as a complex function. It keeps us on the safe side.
\end{rem}
\begin{lem}If $\{a_n\}_{n=0}^\infty$ is a sequence of non-negative real numbers, where each even term is zero, then $\limsup_{n\rightarrow\infty}a_n=\lim_{k\rightarrow\infty}a_{2k+1}$, if the right hand limit exists.
\end{lem}
\begin{proof}This is follows from the definition of $\limsup$ (which can be found, for example, in \cite{8} page 56). \end{proof}
\begin{proof}[Proof of Proposition 3] In the proof of lemma 2, it was noted that the function $\varphi$ is holomorphic on $\mathbb{D}$. Hence, $\varphi$ is representable by power series in $\mathbb{D}$ (see, e.g., \cite{9} page 207, theorem 10.16). This means that for every point $a\in \mathbb{D}$, there is an open disc $D(a;r):=\{z\in\mathbb{C}:\ |z-a|<r\}\subseteq \mathbb{D}$ and a corresponding power series of the form
$$\sum_{n=0}^\infty c_n(z-a)^n$$
which converges to $\varphi(z)$ for all $z\in D(a;r)$ (\cite{9}, page 198). By the Corollary to \cite{9}'s theorem 10.16 (appearing in \cite{9} on page 199), the constants $c_n$ are unique and are in fact given by
$$c_k=\frac{\varphi^{(k)}(a)}{k!}\ \ \ \ \ (k=0,1,2,...).$$
The radius $r$ cannot exceed a certain number $R\in[0,\infty]$ called ``the radius of convergence" for the series. The radius of convergence $R$ is given by ``Hadamard's formula" (\cite{10} page 38)
$$\frac{1}{R}=\limsup_{n\rightarrow \infty}\left|c_n\right|^\frac{1}{n}.$$
The power series $\sum_{n=0}^\infty c_n(z-a)^n$ converges absolutely and uniformly in $\overline{D}(a;r):=\{z\in\mathbb{C}:\ |z-a|\leq r\}$ for every $r<R$ and diverges if $z\notin \overline{D}(a;R)$ (see \cite{9} page 198). 

We are concerned with the power series representation for $\varphi$ centered at $a=0$. By lemma 2, it follows that the power series representation for $\varphi$ centered at $0$ is
\begin{eqnarray}
\sum_{k=0}^\infty\frac{\varphi^{(k)}(0)}{k!}z^k&=&\sum_{n=0}^\infty\frac{-(2n+2)!2\zeta(2n+3)}{(2n+1)!}z^{2n+1}\nonumber \\
&=&\sum_{n=0}^\infty -(2n+2)2\zeta(2n+3)z^{2n+1}\nonumber \\
&=&-4\sum_{n=0}^\infty(n+1)\zeta(2n+3)z^{2n+1}\nonumber.
\end{eqnarray}
The remaining issue is to find the radius of convergence $R$ for this series. By lemmas 2, 3, and Hadamard's formula, it follows that
$$\frac{1}{R}=\limsup_{k\rightarrow\infty}\left|\frac{\varphi^{(k)}(0)}{k!}\right|^{\frac{1}{k}}=\underbrace{\lim_{k\rightarrow \infty}\left(4(k+1)\zeta(2k+3)\right)^{\frac{1}{k}}}_{\textrm{if this limit exists}}.$$
Our claim then, is that $$\lim_{k\rightarrow\infty}\left(4(k+1)\zeta(2k+3)\right)^\frac{1}{k}=1.$$

One has that $\lim_{k\rightarrow\infty}c^\frac{1}{k}=1$ for any positive constant $c$. So $\lim_{k\rightarrow\infty}4^\frac{1}{k}=1$. To see that $\lim_{k\rightarrow\infty}(k+1)^\frac{1}{k}=1$, set $(k+1)^\frac{1}{k}=1+\delta_k$ (at this point we are adapting an argument from page 39 in \cite{10}). Then we can write $k+1=(1+\delta_k)^k$. Note that $\delta_k$ is positive. The binomial theorem gives
\begin{eqnarray}
k+1&=&(1+\delta_k)^k\nonumber  \\
&>& 1+  {k\choose2}\delta_k^2\nonumber \\
&=&1+\frac{1}{2}k(k-1)\delta_k^2\nonumber.
\end{eqnarray}
Therefore, provided $k>1$, we get that $\delta_k^2<\frac{2}{k-1}$. Hence $\delta_k\rightarrow0$ as $k\rightarrow\infty$ and this proves that $\lim_{k\rightarrow\infty}(k+1)^\frac{1}{k}=1$

It remains to be shown that $\lim_{k\rightarrow\infty}\left(\zeta(2k+3)\right)^\frac{1}{k}=1$. Note that it suffices to show that $\lim_{k\rightarrow\infty}\zeta(k)=1$ ($k\in\mathbb{Z}^+$) since it would then follow that $\frac{1}{2}\leq\zeta(2k+3)\leq 2$ for sufficiently large $k$, and hence $\lim_{k\rightarrow\infty}\left(\zeta(2k+3)\right)^\frac{1}{k}=1$ by the ``squeeze theorem" of freshman calculus.

We have that $\zeta(k):=\sum_{n=1}^\infty\frac{1}{n^k}=1+\sum_{n=2}^\infty\frac{1}{n^k}$, and hence $\lim_{k\rightarrow\infty}\zeta(k)=1+\lim_{k\rightarrow\infty}\sum_{n=2}^\infty\frac{1}{n^k}$.

We can pass the limit through the summation symbol, obtaining
$$\lim_{k\rightarrow\infty}\zeta(k)=1+\sum_{n=2}^\infty\lim_{k\rightarrow\infty}\frac{1}{n^k}=1.$$Passing the limit through the summation symbol is justified for the following reason.

Notice that $\frac{1}{n^k}\rightarrow0$ as $k\rightarrow\infty$ (since $n>1$) and $\left|\frac{1}{n^k}\right|\leq 1$ for all $k$. Thus, by the ``Lebesgue Dominated Convergence Theorem" (\cite{9} page 26), it follows that 
$$\lim_{k\rightarrow\infty}\int_X \frac{1}{n^k} d\mu(n)=\int_X\lim_{k\rightarrow\infty}\frac{1}{n^k} d\mu(n),$$ where $X=\{2,3,4,....\}$ and $\mu$ is the counting measure.
In other words, $$\lim_{k\rightarrow\infty}\sum_{n=2}^\infty\frac{1}{n^k}=\int_X 0d\mu =0,$$
as claimed.
\end{proof}
\newpage
\
\\
{\Large\textbf{Acknowledgments}} 

\
\\
The author would like to thank Larry Weaver, Dave Auckly, David Yetter, and especially his doctoral advisor, Louis Crane, for helpful suggestions.

\end{document}